\documentclass[a4paper,twocolumn,10pt, showpacs]{article}

\usepackage[nointlimits]{amsmath}
\usepackage{hyperref}
\usepackage{cite}
\usepackage{amsfonts}
\usepackage{amssymb}
\usepackage{amsmath}
\usepackage{amsthm}
\usepackage{latexsym}
\usepackage{graphicx}
\usepackage{color}
\usepackage{layout}
\usepackage[english]{babel}
\usepackage[utf8]{inputenc}

\allowdisplaybreaks

\usepackage{geometry}
\begin{document}

\title{Inflation with Planck: a survey of some ``exotic" inflationary models}

\author{
	Cláudio Gomes$^{1}$\footnote{E-mail: claudio.gomes@fc.up.pt}, Orfeu Bertolami$^{1}$\footnote{E-mail: orfeu.bertolami@fc.up.pt}, João G. Rosa$^{2}$\footnote{E-mail: joao.rosa@ua.pt}\\
	\\
	\normalsize{$^{1}$Departamento de Física e Astronomia, Faculdade de Ciências da Universidade do Porto} \\
	\normalsize{and Centro de Física do Porto,}\\
	\normalsize{Rua do Campo Alegre 687, 4169-007 Porto, Portugal}\\
	\\
	\normalsize{$^{2}$ Departamento de F\'isica da Universidade de Aveiro and CIDMA,}\\
	\normalsize{Campus de Santiago, 3810-183, Aveiro, Portugal}\\
	\\
}



\date{\today}
\maketitle
\begin{abstract}
We examine some inflationary models based on modifications of gravity in the light of Planck 2015 data, such as the generalised Chaplygin inspired inflation, models based in $N=1$ supergravity and braneworld scenarios. We also show that, conversely, potentials with a very flat plateau yield a primordial spectrum similar to that of the Starobinsky model with no need to modify general relativity. 
\end{abstract}


\section{Introduction}

Inflation is a well established paradigm in Cosmology. It solves the initial conditions problems of the standard Hot Big Bang model, namely homogeneity, isotropy, flatness of the Universe and the overabundance of magnetic monopoles putatively generated in the context of Grand Unified Theories. It also provides a mechanism to understand the origin of the observable large scale structures due to quantum fluctuations of the inflaton field. There are many models in which this paradigm can be achieved. Therefore, over the past decades several space-based observatories have gathered cosmic information and posed severe restrictions on sets of cosmological parameters as the COBE \cite{cobe}, the WMAP \cite{wmap} and the Planck satellites \cite{planck}.

Starobinsky's model \cite{starobinsky} was the first proposal to solve the above problems from a modification of gravity for high curvatures, and can also be written in terms of a scalar field in the Einstein frame. This model is compatible with Planck data at $1\sigma$ level. Afterwards, other scalar field models were proposed in the context of general relativity, as the old and new inflation scenarios \cite{linde,guth,steinhardt}. The first realizations of inflation are excluded on various grounds.

More recently, several inflationary scenarios were constructed in the context of alternative theories of gravity, such as $f(R)$ models (see Ref. \cite{felice} for a review). Some extensions of these theories including non-minimal inflaton-curvature couplings were also analysed, being compatible with Planck's data for some inflationary potentials \cite{nmc,odintsov,kaiser,nmcinf}.

There are several models of cosmic inflation which are based on scalar fields from superstring and supergravity theories \cite{olive}. One of such cases is the so called D-brane inflation \cite{Burgess:2001fx}, where the inflaton is a geometric modulus corresponding to the distance between two stacks of three-dimensional branes and anti-branes, with the Standard Model particles being confined to the branes that survive brane-antibrane annihilation at the end of inflation. The quadratic \cite{bellido} and quartic \cite{string,dvali} models are compatible with Planck's results \cite{planck}. 

Another possibility is assuming that inflation can be driven by the Higgs field non-minimally coupled with the scalar curvature \cite{shaposhnikov}. This is compatible with data \cite{shaposhnikov2}.

Furthermore, there are other inflationary models with more than one scalar field. An example of that is the so-called hybrid inflation \cite{hybrid0,hybrid}, where a rapid rolling, or waterfall, of a scalar field, $\sigma$, is prompted by a second scalar field, $\phi$. However, these models predict a blue tilted power spectrum and are therefore ruled out.

An alternative realization of the inflationary paradigm is warm inflation, where the inflaton is coupled to other light fields \cite{warm,warm2,warm3,warm4}. As the inflaton slow-rolls along the potential, it loses energy that is transferred to radiation, thus heating the Universe without the need of a reheating stage after the slow-roll period.

Aside from scalar field inflation, there are also a few models on inflation driven by vector fields \cite{vector,vector2}. This is hard to achieve unless higher curvature terms are taken into consideration, thus allowing for an exponential expansion of the scale factor and for attractor points \cite{vector3}. However, further analysis needs to be done in order to fully compare these models with existing data.

Inflation is thus a prime arena for exploring physical theories that go beyond general relativity and the Standard Model of particle physics, with the above discussion being just a modest sample of the large body of literature devoted to this subject. There are, however, several non-standard inflationary models, in particular those involving modifications of general relativity, for which observational predictions have yet to be compared with the recent Planck data, a gap that we wish to fill in with this work. In  Ref. \cite{nmcinf}, we have developed a numerical code to explore inflationary scenarios for a generalised Friedmann equation $H^2 = H^2 (\rho)$, and we will employ this tool to explore the generalised Chaplygin gas inflationary model, as well as some inflationary models within supergravity and braneworld scenarios leading to modifications of the Friedmann equation. We will show that observations allow for significant deviations from general relativity in the context of such models.

Conversely, we also show, at the end of our discussion, that general relativity can yield observational predictions that are degenerate with some modified gravity models, considering in particular the case of plateau-like potentials that yield a primordial spectrum of fluctuations similar to that of the Starobinsky model mentioned above.

This work is organised as follows. In Section \ref{sec:chaplygin}, we analyse the monomial and hilltop potentials in the context of the Chaplygin inspired inflation. In Section \ref{sec:susy}, a model of $N=1$ Supergravity inflation is examined. In Section \ref{sec:sgb}, the monomials and hilltop inflationary potentials are studied for supergravity inflation in the context of the braneworld scenario. In Section \ref{sec:n1sgb}, we consider the case of $N = 1$ supergravity on the brane for different potentials. In Section \ref{sec:plateau}, we explore the predictions for nearly-flat plateaux within general relativity. Finally, we summarize our main conclusions in Section \ref{sec:conclusions}.

\section{Chaplygin-inspired inflation}\label{sec:chaplygin}
The generalised Chaplygin model unifies dark matter and dark energy adopting an exotic fluid with equation of state \cite{chaplygin,pasquier}:
\begin{equation}
p_{GCG}=-\frac{A}{\rho^{\alpha}_{GCG}} ~,
\end{equation}
where $p_{GCG}$ and $\rho_{GCG}$ are the pressure and the energy density of the gas, respectively, and $A$ and $\alpha$ are constants, with $0 < \alpha \leq 1$. This unified model can be built with an underlying scalar field with a suitable potential \cite{chaplygin} \footnote{Notice that the model also admits a complex scalar field construction \cite{chaplygin, chapcomplex}.}.

Such an equation of state gives rise to a fluid energy density that evolves as $\rho_{GCG}=\left(A+\rho_m^{1+\alpha}\right)^{\frac{1}{1+\alpha}}$, where $\rho_m\propto a^{-3}$ behaves as the energy density of non-relativistic matter. It has been shown in Ref. \cite{chapinf} that such a behaviour can also be obtained from a modification of gravity, in particular from a generalised Born-Infeld action for a scalar field, yielding a Friedmann equation of the form:
\begin{equation}
H^2=\frac{1}{3M_P^2}\left[A+\rho_{\phi}^{1+\alpha}\right]^{\frac{1}{1+\alpha}}~,
\end{equation}
where $M_P= (8\pi G)^{-2}$ is the reduced Planck mass. One can therefore assume that inflaton is such a scalar field, nevertheless obeying the standard classical equation of motion:
\begin{equation}
\ddot{\phi}+3H\dot{\phi}+V'(\phi)=0~.
\label{fieldeq}
\end{equation}

As examples, we will consider two main cases in the slow-roll regime: $\alpha=1$ and $\alpha=0.5$ for linear and quadratic monomials inflationary potentials. However, for the quartic and quadratic hilltop models only the $\alpha=1$ case will be analysed due to numeric limitations in the computation of the observable quantities. Monomial potentials are the ones which can be cast in the following form:
\begin{equation}
V(\phi)=V_0\left(\frac{\phi}{M_P}\right)^n~,
\end{equation}
where the constant $V_0$ sets the scale of inflation and $n \in \mathbb{Z}^+$. For these potentials, the slow-roll parameters read:
\begin{eqnarray}
\epsilon_{\phi}&=&\frac{n^2}{2\tilde{\phi}^2}~,\nonumber\\
\eta_{\phi}&=&\frac{n(n-1)}{\tilde{\phi}^2} ~,
\end{eqnarray}
where $\tilde{\phi}\equiv \phi/M_P$.

Hilltop potentials can be written as:
\begin{equation}
V(\phi)=V_0\left(1-\frac{\gamma}{n}\tilde{\phi}^n\right) ~,
\end{equation}
where the parameter $\gamma \in ]0, 1[$. The slow-roll parameters can be expressed as:
\begin{eqnarray}
\epsilon_{\phi}&=&\frac{\gamma^2\tilde{\phi}^{2n-2}}{2\left(1-\frac{\gamma}{n}\tilde{\phi}^n\right)^2}~,\nonumber\\
\eta_{\phi}&=&-\frac{(n-1)\tilde{\phi}^{n-2}}{1-\frac{\gamma}{n}\tilde{\phi}^n} ~.
\end{eqnarray}

The number of e-folds of inflation after horizon-crossing reads:
\begin{eqnarray}
N_e &=& -\frac{1}{M_P^2}\int_{\phi_*}^{\phi_e} \frac{\left(A + V(\phi)^{1 + \alpha}\right)^{\frac{1}{1 + \alpha}}}{V'(\phi)} d\phi ~,
\end{eqnarray}
where the value of the inflaton at the end of inflation for a generic modified Friedmann equation of the type $H^2=H^2(\rho)$, is given by the strongest of the two conditions for each potential \cite{nmcinf}:
\begin{eqnarray}
\epsilon_{\phi}&\sim  & 3\left(\frac{M_P^2H^2}{V}\right)^2\left(M_P^2\frac{dH^2}{d\rho}\right)^{-1}~,\\
\epsilon_{\phi} &\sim & 9\frac{H^2M_P^2}{V}~,
\end{eqnarray}
where the first condition signals the end of accelerated expansion and the second condition the failure of the slow-roll condition $\dot\phi^2/2 < V(\phi)$. Note that these two conditions coincide in general relativity with a canonical scalar field, but are in general distinct when one considers deviations from the standard Friedmann equation, as is the case of the generalised Chaplygin inflation scenario. 

The scalar spectral index and the tensor-to-scalar ratio are given by:
\begin{eqnarray}
n_s &=& 1 - 6 \epsilon_{\phi_*} \frac{V(\phi_*)^{2 + \alpha}}{\left(A + V(\phi_*)^{1 + \alpha}\right)^{\frac{\alpha + 2}{\alpha + 1}}}  +\\
&& + \frac{2 \eta_{\phi_*}V(\phi_*)}{\left(A + V(\phi_*)^{1 + \alpha}\right)^{\frac{1}{1 + \alpha}}}~,\nonumber\\
r&=&16 \epsilon_{\phi_*} \frac{V(\phi_*)^2}{\left(A + V(\phi_*)^{1 + \alpha}\right)^{\frac{2}{1 + \alpha}}}~.
\end{eqnarray}

\begin{figure}[h!]
\centering
\includegraphics[width=\columnwidth]{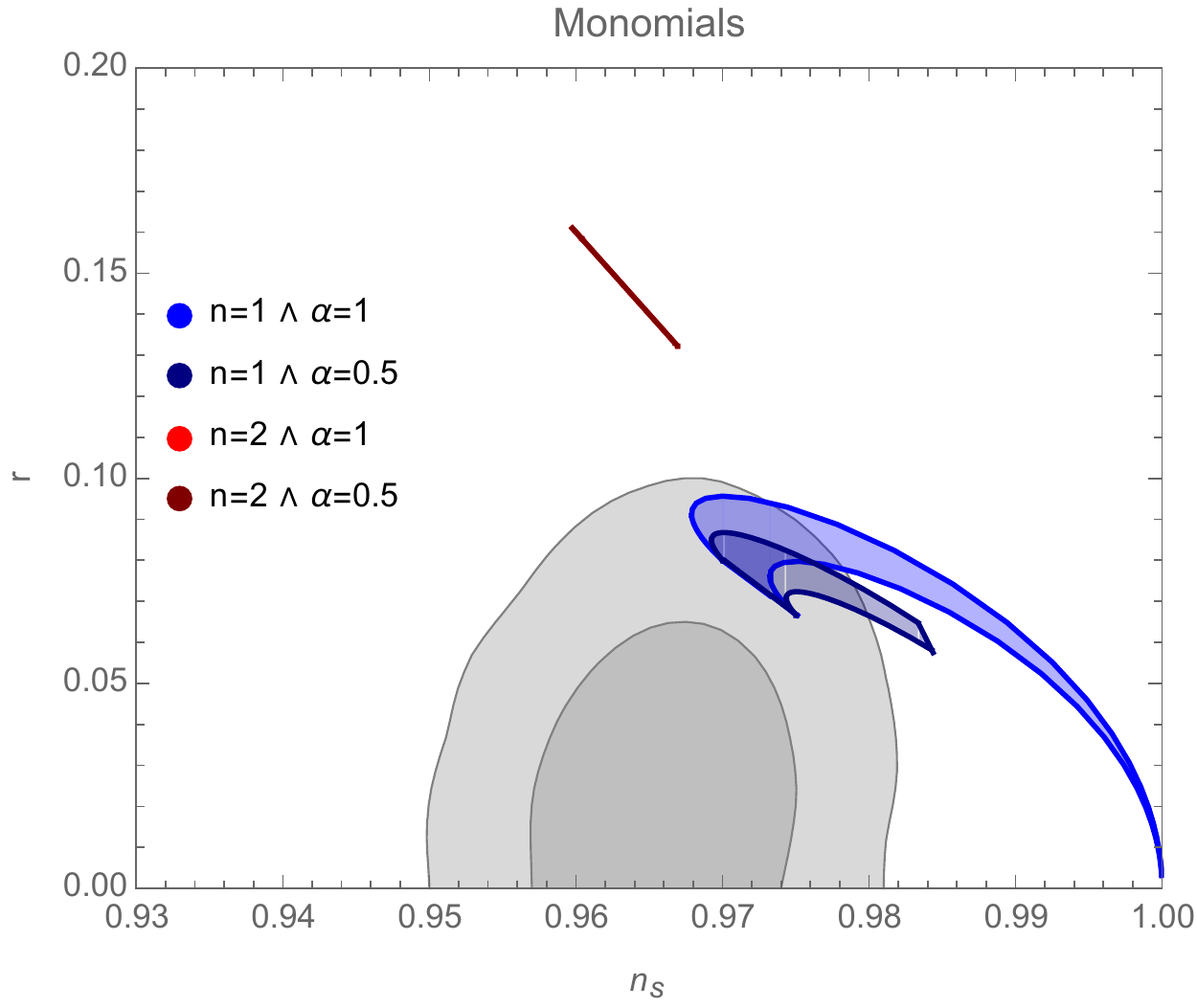}
\caption{Predictions for monomial potentials in the Chaplygin inspired inflation in comparison with Planck data in grey. For $n = 1$, if $\alpha = 0.5$, then $x\equiv \frac{A}{V_0^{1+\alpha}}< 0.53$. For $n = 2$, if $\alpha=1$, then $x < 0.59$, and for $\alpha = 0.5$ we
must have $x<0.53$. Upper and lower bounds correspond to $N_e = 50$ and $N_e = 60$, respectively.}
\label{fig:chapmono}
\end{figure}

Let us define $x \equiv \frac{A}{V_0^{1+\alpha}}$. In Fig. \ref{fig:chapmono}, we plot the observational predictions for the linear and quadratic potentials. From the condition that inflation has to end, some constraints arise for $x$ in all cases, except for the linear potential for $\alpha=1$, for which it is always possible to end inflation without further requirements on $x$ and that exhibits an attractor point at $(n_s,r)=(1,0)$ for $x \to \infty$.

\begin{figure}[h!]
\centering
\includegraphics[width=\columnwidth]{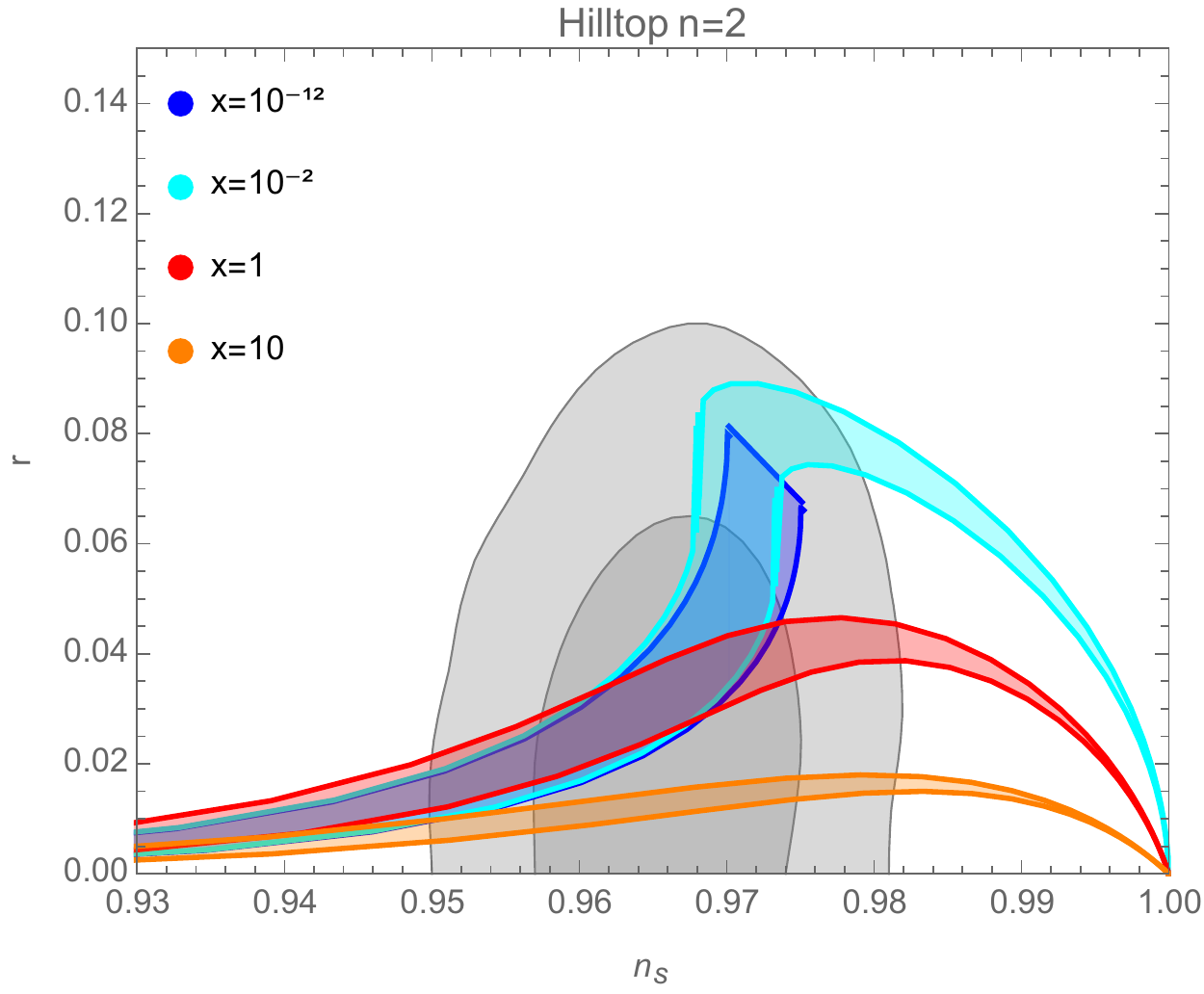}
\caption{Predictions for the quadratic hilltop potential in a generalised Chaplygin model with $\alpha = 1$ in comparison with Planck data in grey. For $x\ll 1$, one retrieves the usual Friedmann equation prediction, whilst for larger values of $x$ the potential behaviour differs from that. Upper and lower bounds correspond to $N_e = 50$ and $N_e = 60$, respectively.}
\label{fig:chaphill2}
\end{figure}

\begin{figure}[h!]
\centering
\includegraphics[width=\columnwidth]{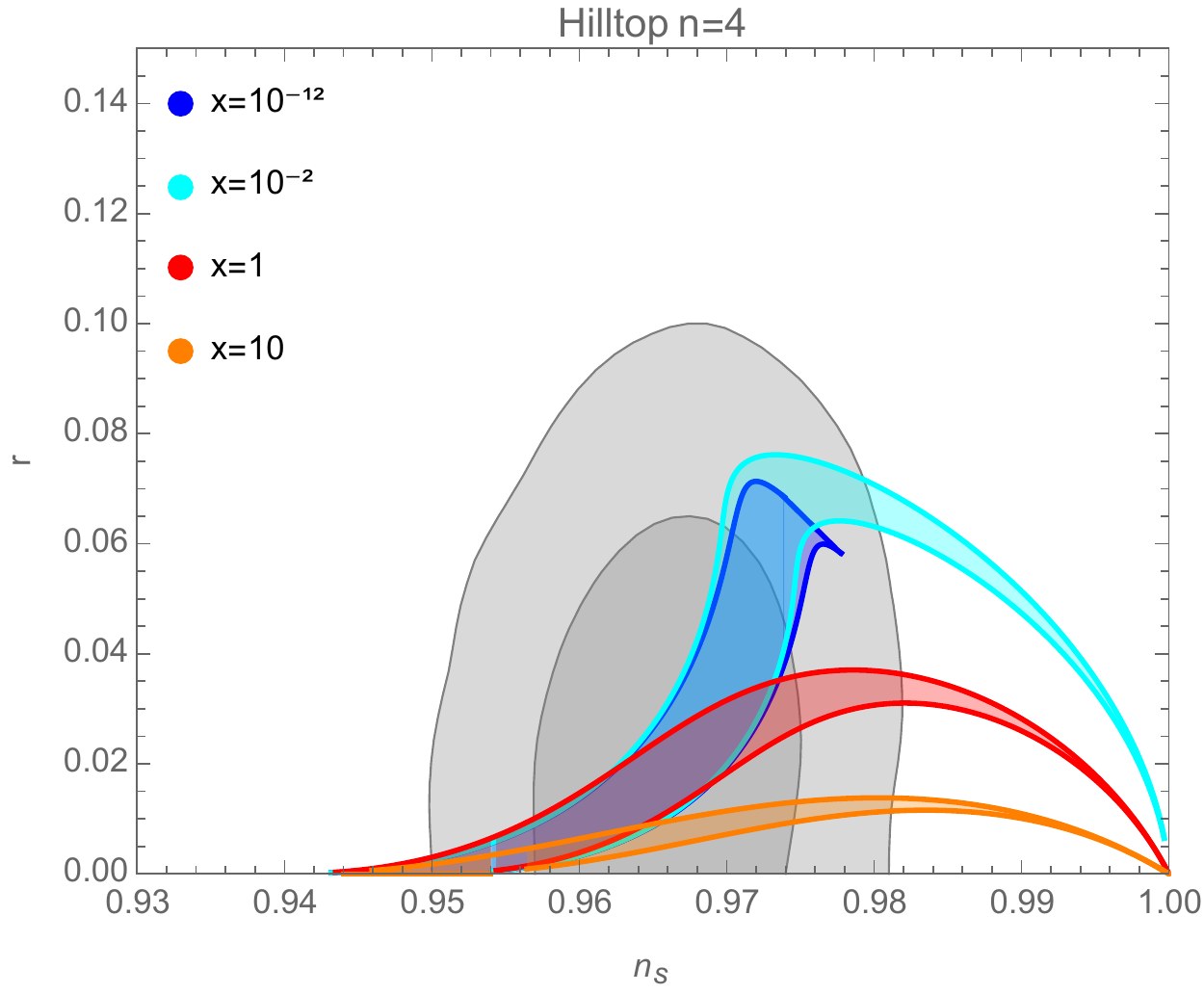}
\caption{Predictions for the quartic hilltop potential in a generalised Chaplygin model with $\alpha = 1$ in comparison with Planck data in grey. For $x\ll 1$, one retrieves the usual Friedmann equation prediction, whilst for larger values of $x$ the potential behaviour is quite different. Upper and lower bounds for each case correspond to e-folds between $50$ and $60$, respectively.}
\label{fig:chaphill4}
\end{figure}

For $\alpha=1$, the observational predictions for the quadratic and quartic hilltop potentials are shown in Fig. \ref{fig:chaphill2} and Fig. \ref{fig:chaphill4}, respectively, for different values of $x$. In both cases, as $x\to 0$, one retrieves the predictions of such potentials in general relativity, as expected, but for greater $x$ the deviations become significant: the tensor-to-scalar ratio becomes smaller and there is an attractor point at $(n_s,r)=(1,0)$ for $\gamma\to 0$. However, there is still a range of values for the free parameter that is allowed by data.

\section{$N=1$ Supergravity Inflation}\label{sec:susy}
We shall now study an $N=1$ supergravity (SUGRA) potential that describes the interaction of chiral superfields, and is specified by the K\"ahler potential, $K(\Phi,\Phi^{\dagger})$. The scalar potential reads \cite{susy}:
\begin{equation}
V=\frac{e^K}{4}\left(G_a(K^{-1})^a_bG^b-3|W|^2\right) ~,
\end{equation}
where $G_a = K_a W +W_a$ is the Kähler function, the indices $a,~b$ correspond to derivatives with respect to the chiral superfields, $\Phi$, and $W(\Phi)$ is the superpotential that describes the Yukawa and the scalar couplings of the supersymmetric theory.

We shall assume that the superpotential can be split into supersymmetry-breaking, $P$, gauge, $G$, and inflationary, $I$, sectors, such that:
\begin{equation}
W=P+G+I~.
\end{equation}

We will consider the supersymmetric model constructed in Refs. \cite{ross0,ross} with the minimal choice for the K\"ahler potential, $K =\Phi\Phi^{\dagger}$ and the inflaton superpotential, $I = \Delta^2 M_P f (\frac{\Phi}{M_P})$, where $\Delta$ corresponds to the inflation scale, and $f (\Phi/M_P )$ is a function that is not constrained by the underlying $R$-symmetry of the model. Thus, the scalar potential in terms of the inflaton field, $\phi$, reads \cite{susy}:
\begin{equation}
V_I(\phi)=e^{|\phi |^2/M_P^2}\left(\left|\frac{\partial I}{\partial \phi}+\frac{\phi^* I}{M_P^2}\right|^2-\frac{3|I|^2}{M_P^2}\right)_{\Phi=\phi} ~.
\end{equation}

Requiring that SUSY remains unbroken in the global minimum, $\left|\frac{\partial I}{\partial \Phi}+\frac{\Phi^*I}{M_P^2} \right|_{\Phi=\phi_0}=0$, and that the present cosmological constant vanishes, $V_I(\phi_0)=0$, it follows that the simplest superpotential $I$ that satisfies these conditions is:
\begin{equation} 
I(\phi)=V_0(\phi-\phi_0)^2 ~,
\end{equation} 
where $V_0=\Delta^2/M_P$. This potential is flat close to the origin for $\phi_0=M_P$, which also yields a vanishing inflaton mass at this point. Thus, in the vicinity of the origin the potential is given by the Taylor expansion \cite{ross0,ross}:
\begin{equation}
V_I(\phi)=\Delta^4\left(1-4\tilde\phi^3+\frac{13}{2}\phi^4-8\tilde\phi^5+\frac{23}{3}\tilde\phi^6+\dots\right) ~.
\end{equation}

In this model, the slow-roll parameters read:
\begin{eqnarray}
\epsilon_{\phi}&\simeq& 2 \left(\frac{-6 \tilde{\phi} ^2 + 13 \tilde{\phi} ^3 - 20 \tilde{\phi} ^4 + 23 \tilde{\phi} ^5 }{
   1 - 4 \tilde{\phi} ^3 + \frac{13}{2} \tilde{\phi} ^4 - 8 \tilde{\phi} ^5 + \frac{23}{3} \tilde{\phi} ^6}\right)^2 ~,\nonumber\\
\eta_{\phi}&\simeq&\frac{-24 \tilde{\phi} + 78 \tilde{\phi} ^2 - 160 \tilde{\phi}^3 + 230 \tilde{\phi} ^4}{1 -  4 \tilde{\phi} ^3 + \frac{13}{2} \tilde{\phi} ^4 - 8 \tilde{\phi} ^5 + \frac{23}{3} \tilde{\phi} ^6}~.
\end{eqnarray}

The number of e-folds follows from the integration:
\begin{eqnarray}
N_e= -{1\over M_P^2}\int_{\phi_*}^{\phi_e} {V(\phi)\over V'(\phi)}d\phi ~. 
\end{eqnarray}

In such models, the equation of motion of the inflaton and the Friedmann equation are the standard ones in general relativity. Therefore, the scalar index and the tensor-to-scalar ratio are straightforwardly computed at horizon crossing:
\begin{eqnarray}
n_s &=& 1-6\epsilon_{\phi_*}+2\eta_{\phi_*}~,\\
r &=& 16\epsilon_{\phi_*}~.
\end{eqnarray}

This model is ruled out, since it predicts a scalar index in the range $n_s \in [0.92,0.94]$ which is excluded by the Planck data. It also leads to very small tensor-to-scalar ratio, actually of the order of $10^{-9} - 10^{-8}$. Since this SUGRA model does not meet observational data, we shall consider supergravity inflation on the brane in the next sections.

\section{Brane Inflation}\label{sec:sgb}

In the 5-dimensional brane scenario, matter fields are confined to a 3-brane and only gravity propagates in the fifth dimension, and the 4-dimensional Einstein equations read \cite{sgb}:
\begin{equation}
G_{\mu\nu}=-\Lambda g_{\mu\nu}+\frac{1}{M_P^2}T_{\mu\nu}+\frac{1}{M_5^6}S_{\mu\nu}-E_{\mu\nu}~,
\end{equation}
where $M_5 = (8\pi G_5 )^{-2}$ is the 5-dimensional reduced Planck mass associated to the 5-dimensional gravitational constant $G_5$, $T_{\mu\nu}$ is the energy-momentum on the brane, $S_{\mu\nu}$ is a tensor with contributions quadratic in $T_{\mu\nu}$, and $E_{\mu\nu}$ is the projection of the 5-dimensional Weyl tensor on the 3-brane. We assume that the brane is described by the Robertson-Walker metric, such that the Friedmann equation reads \cite{sgb,sgb2,sgb3}:
\begin{equation}
H^2=\frac{\Lambda}{3}+\frac{\rho}{3M_P^2}+\frac{\rho^2}{9M_5^6}+\frac{\epsilon}{a^4}~,
\end{equation}
where $\epsilon$ is an integration constant and $a$ is the scale factor. Since the cosmological constant is negligible and the last term of the Friedmann equation quickly vanishes when inflation sets in, the modified Friedmann equation reads \cite{sgb4,n1sgb}:
\begin{equation}
H^2=\frac{\rho}{3M_P^2}\left(1+\frac{\rho}{2\lambda}\right) ~,
\end{equation}
where $\lambda$ is the 3-brane tension \footnote{We note that there are other similar models as the Cardassian one \cite{freese}, where the Friedmann equation has an {\it ad hoc} correction of the form $\left(1+\frac{\rho^n}{2\lambda}\right)$; this model has already been constrained by several observations, and both the $\Lambda$CDM and the
Cardassian models are equally favoured with WMAP \cite{sen}, whilst Gamma Ray Burst data further narrow the allowed values for the exponent of such model or even a modified version of it - the modified polytropic Cardassian model \cite{cardassiangrb}. Another possibility is loop quantum cosmology \cite{lqc,lqc2}, where the corrections to the standard Friedman equation have a minus sign, $H^2 \sim \rho \left(1-\frac{\rho}{2\lambda}\right)$, which is compatible with data \cite{lqcinflation,bouncelqc}.}.

This model does not change the equation of motion for the fields, namely Eq. (\ref{fieldeq}) (see, however, Ref.~\cite{carvalho}).

In order to achieve a successful inflationary period, one must require the slow-roll parameters to be small. The number of e-folds of inflation follows from:
\begin{eqnarray}
N_e=-\int \frac{1}{\sqrt{2 \epsilon_{\tilde{\phi}}} }\left( 1 + \frac{1}{2\lambda} V(\tilde{\phi})\right) d\tilde{\phi} ~.
\end{eqnarray}

Finally, the scalar index and the tensor-to-scalar ratio can be written as:
\begin{eqnarray}
n_s &=&  1 - 6 \epsilon_{\phi_*} \frac{1 + \frac{1}{\lambda} V(\phi_*)}{\left(1 + \frac{1}{2\lambda} V(\phi_*)\right)^2} +  \frac{2\eta_{\phi_*}}{\left(1 + \frac{1}{2\lambda} V(\phi_*)\right)^2}~.\nonumber\\
r &=& 16 \epsilon_{\phi_*} \frac{1}{\left(1 + \frac{1}{2\lambda} V(\phi_*)\right)^2}~.
\end{eqnarray}

\begin{figure}[h!]
\centering
\includegraphics[width=\columnwidth]{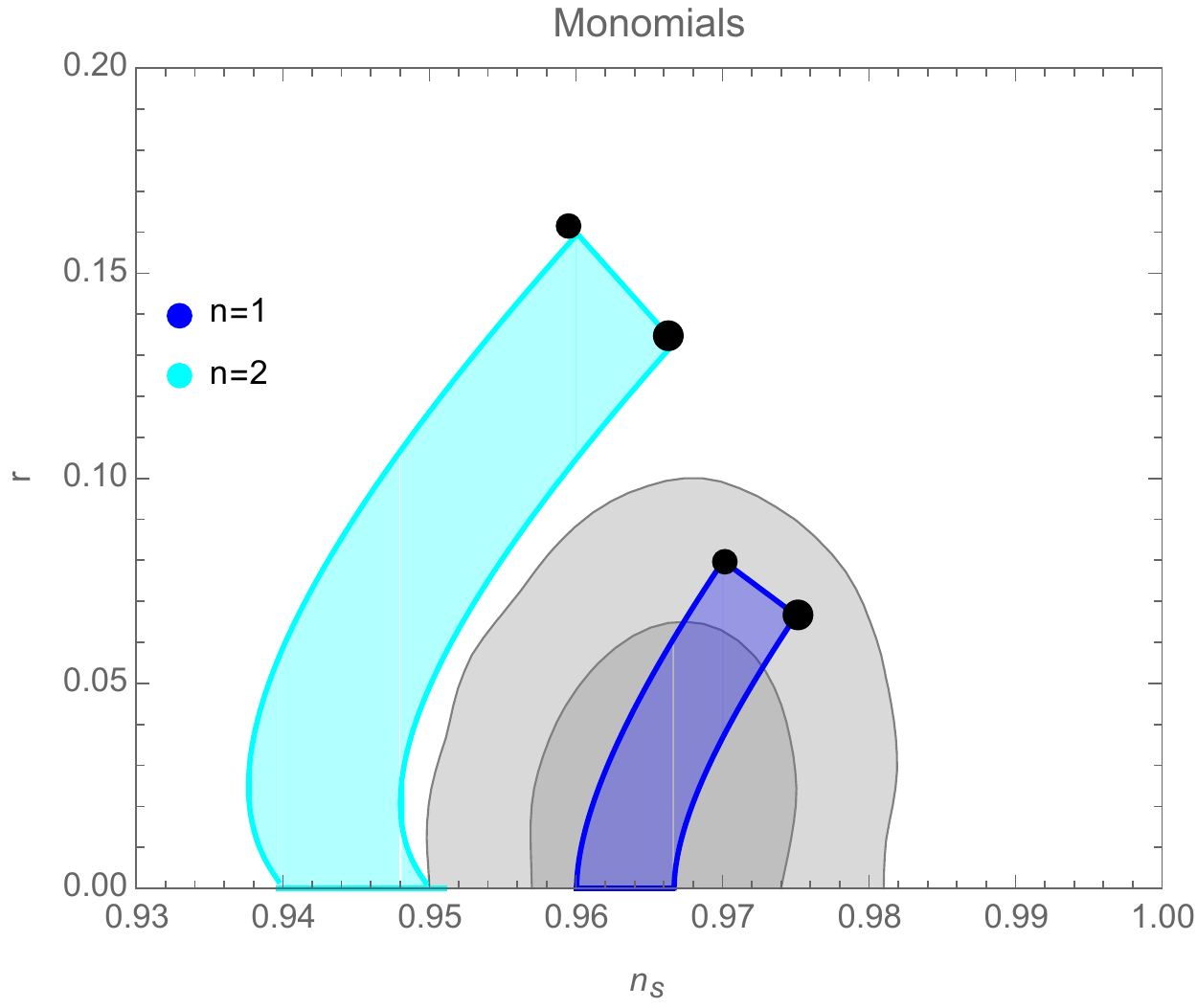}
\caption{Predictions for monomial potentials in a braneworld inflation scenario. For the factor $y \equiv \frac{V_0}{\lambda} \sim 0$, one approaches the linear or the quadratic behaviour in the standard Friedmann equation, represented by the black dots, but as $y$ grows, the deviation predicts lower tensor-to-scalar ratios. The quadratic model is in conflict with Planck observations, but the linear potential is always compatible with data, and for $y\simeq 0.05$ it becomes $1\sigma$ compatible with Planck data (in grey). Upper and lower bounds for each potential correspond to $N_e = 50$ and $N_e = 60$, respectively.}
\label{fig:sgbmono}
\end{figure}

For linear and quadratic monomial potentials, observational predictions are plotted in Fig. \ref{fig:sgbmono}. The quadratic monomial model is ruled out by Planck data, but the linear model is in good agreement. When the ratio $y\equiv \frac{V_0}{\lambda}$ grows, the tensor-to-scalar ratio gets smaller and, in particular, for the ratio $y \geq 0.05$ the linear potential becomes $1\sigma$ compatible with data.

\begin{figure}[h!]
\centering
\includegraphics[width=\columnwidth]{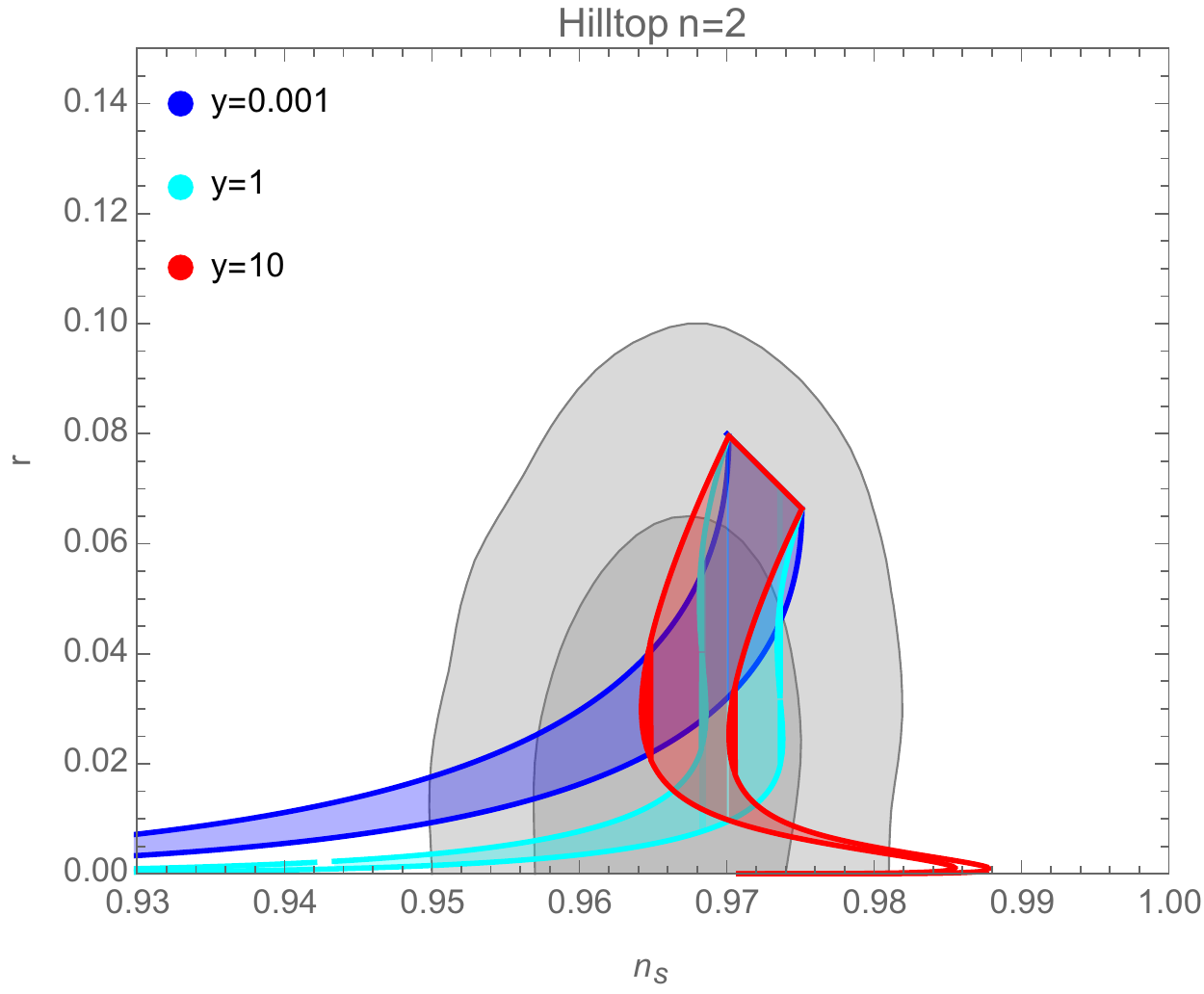}
\caption{Predictions for the quadratic hilltop potential in a braneworld inflationary scenario in comparison with Planck data in grey. When the factor  $y \equiv \frac{V_0}{\lambda} \sim 0$, one gets the linear behaviour with the standard Friedmann equation, but as $y$ grows, one predicts larger values of $n_s$ for $\gamma\to 1$. Nevertheless, they are compatible with data for a large subset of the parameter space. Upper and lower bounds correspond to $N_e = 50$ and $N_e = 60$, respectively.}
\label{fig:sgbhill2}
\end{figure}

\begin{figure}[h!]
\centering
\includegraphics[width=\columnwidth]{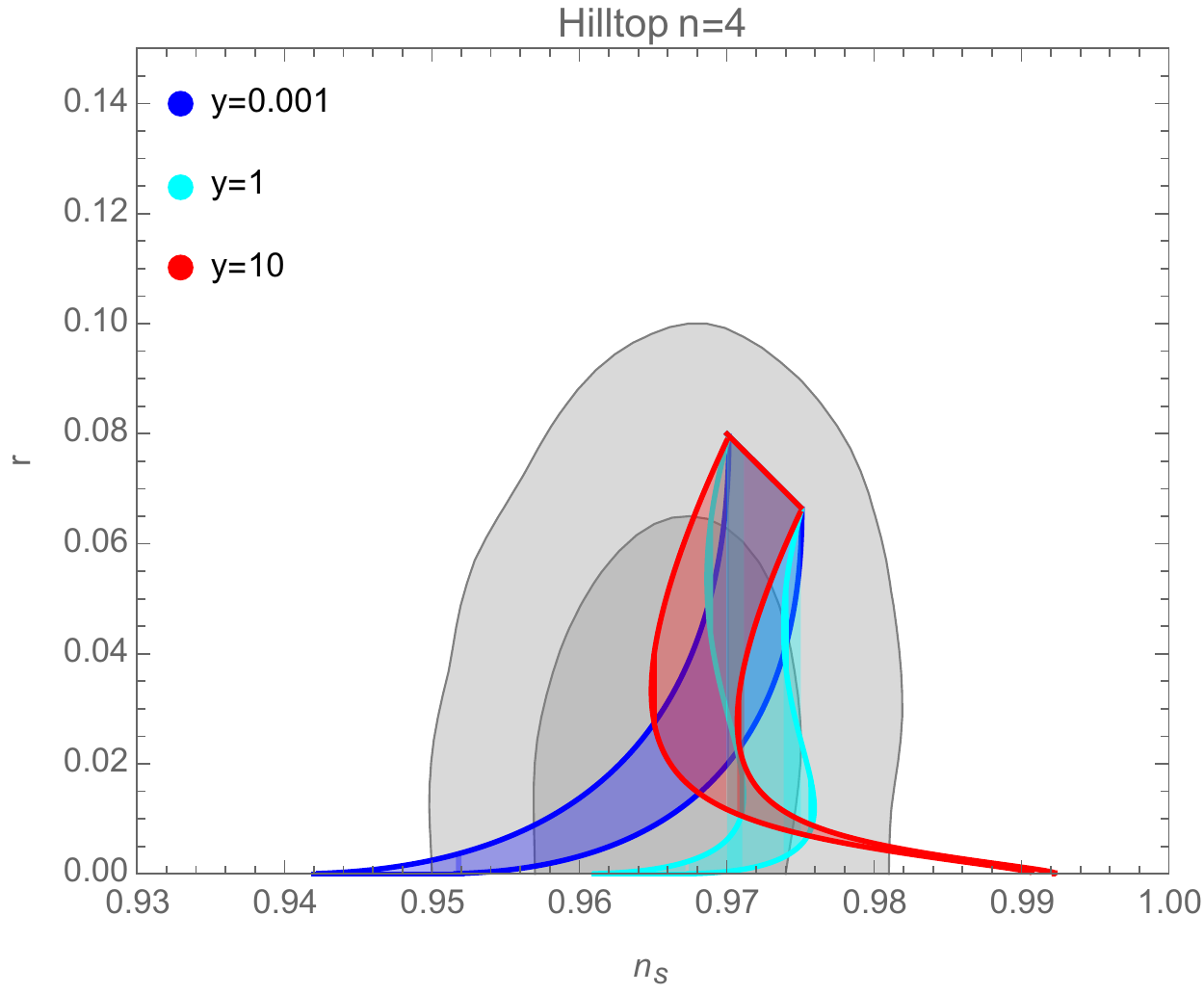}
\caption{Predictions for the quartic hilltop potential in a braneworld inflationary scenario in comparison with Planck data in grey. When the factor  $y \equiv \frac{V_0}{\lambda} \sim 0$, one gets the linear potential limit with the standard Friedmann equation, but as $y$ grows, the deviation leads to larger values of $n_s$ for $\gamma\to 1$. Nevertheless, they are compatible with data for a large subset of the parameter space. Upper and lower bounds for each case correspond to e-folds between $50$ and $60$, respectively.}
\label{fig:sgbhill4}
\end{figure}

The quadratic and hilltop potentials are analysed in Fig. \ref{fig:sgbhill2} and \ref{fig:sgbhill4}, respectively. As the $y$ factor grows, the predictions tend to larger values of the scalar spectral index. Notwithstanding, taking the limit $\gamma\to 0$, we recover the general relativity predictions for a linear potential, regardless of the value of the parameter $y$.

\section{$N=1$ Supergravity inflation in the braneworld scenario}\label{sec:n1sgb}

Let us consider the braneworld scenario discussed in the previous section, now in $N=1$ SUGRA as in Sec. \ref{sec:susy}. However, instead of considering the case $\phi_0=M_P$ that leads to a hilltop-like potential, let us consider the case $\phi_0=0$, which yields a large-field inflationary model and consequently more significant deviations from the standard Friedmann equation.

The relevant part of the inflaton potential (along the real $\phi$ direction) then reads \cite{n1sgb}:
\begin{equation}
V(\phi)=V_0 e^{\tilde{\phi}^2}\left(4\tilde{\phi}^2+\tilde{\phi}^4+\tilde{\phi}^6\right) ~,
\label{potential}
\end{equation}
where, as before, $\tilde{\phi}\equiv \frac{\phi}{M_P}$.
The slow-roll parameters for this potential are given by:
\begin{eqnarray}
\epsilon_\phi&=& 2 \left(\frac{4 + 6 \tilde{\phi} ^2 + 4 \tilde{\phi} ^4 + \tilde{\phi} ^6 }{
   4 \tilde{\phi} + \tilde{\phi}^3 + \tilde{\phi} ^5}\right)^2 ~, \nonumber \\
\eta_\phi &=& \frac{8 + 52 \tilde{\phi} ^2 + 64 \tilde{\phi} ^4 + 30 \tilde{\phi} ^6 + 4 \tilde{\phi} ^8}{4 \tilde{\phi} ^2 + \tilde{\phi} ^4 + \tilde{\phi} ^6} ~,
\end{eqnarray}
and the number of e-folds of the inflation after horizon-crossing is computed from:
\begin{eqnarray}
N_e=-\int \frac{1}{\sqrt{2 \epsilon_{\tilde{\phi}}} }\left( 1 + \frac{1}{2\lambda} V(\tilde{\phi})\right) d\tilde{\phi} ~.
\end{eqnarray}

\begin{figure}[h!]
\centering
\includegraphics[width=\columnwidth]{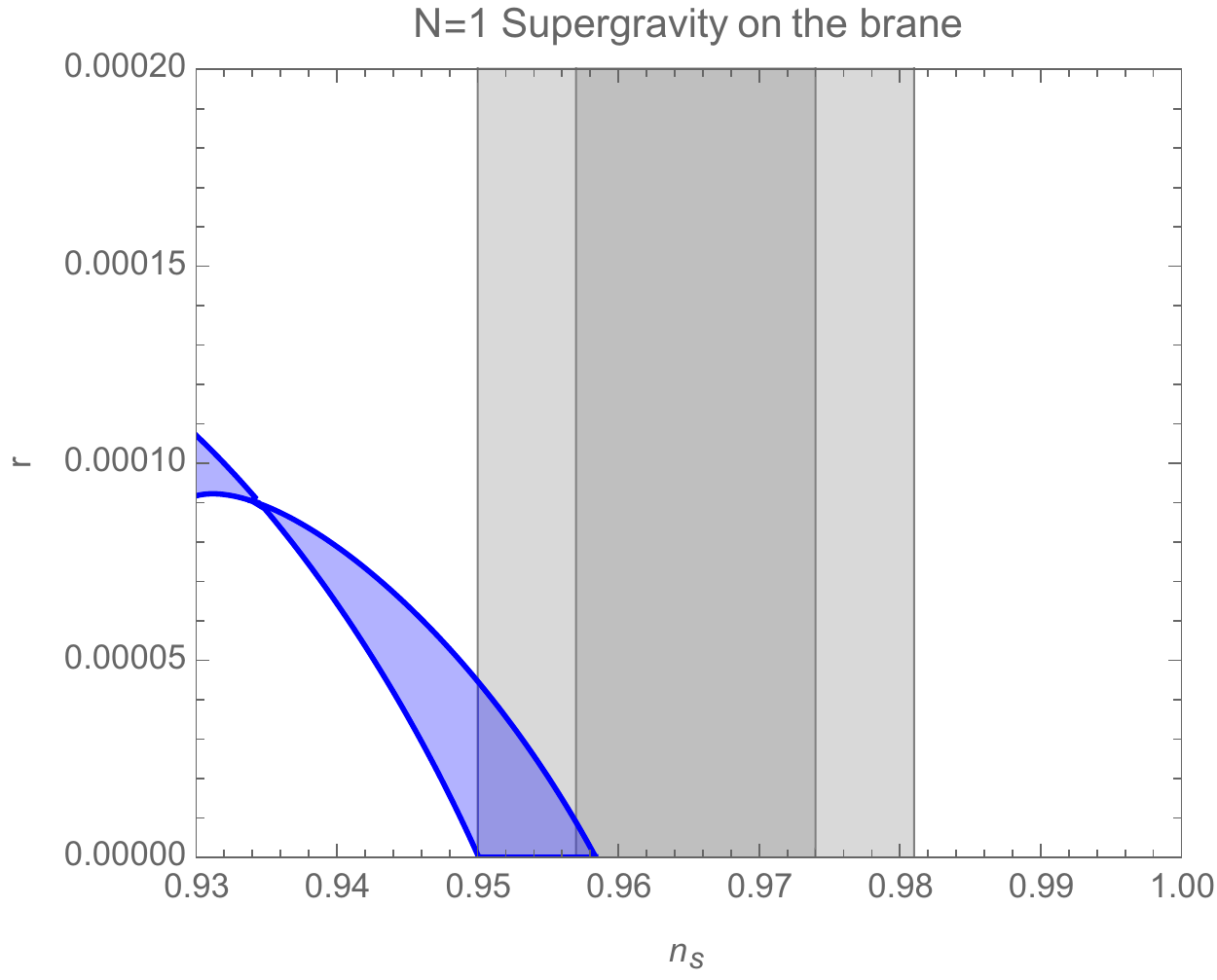}
\caption{Predictions for the $N=1$ supergravity inflation model on the brane. When the factor $y \equiv \frac{V_0}{\lambda}$ increases, predictions go from the left to the right. The upper line at $n_s = 0.93$ corresponds to 50 e-folds of inflation, and the lower one to 60. For 50 e-folds and $y\gtrsim 10^6$ and for 60 e-folds and $y \gtrsim 10^3$ the model becomes compatible with Planck data at 2-$\sigma$ (in light grey).}
\label{fig:susy}
\end{figure}

This potential predicts a very small tensor-to-scalar ratio as shown in Fig. \ref{fig:susy}, being compatible with Planck data within the $2\sigma$ region for  $y \equiv \frac{V_0}{\lambda} \gtrsim 10^6$ for $N_e=50$, and $y \gtrsim 2\times 10^{3}$ for $N_e=60$. 

\section{Plateau potentials in general relativity}\label{sec:plateau}
An inflationary model that has become extremely popular in the recent literature is the Starobinsky model, where a de Sitter phase is driven by an $R^2$ correction to the Einsten-Hilbert action. This model is related to Higgs-inflation and other similar scenarios with a common feature, namely a very flat and negatively curved effective scalar potential in the Einstein frame. The first data release by the Planck collaboration made it very clear that observations favoured potentials with such properties, such as hill-top or natural inflation models in addition to the ones involving modifications of the gravitational sector, at least in the context of single field cold inflation\footnote{In warm inflation, on the other hand, monomial potentials such as $V(\phi)=\lambda\phi^4$ are compatible with Planck data \cite{warm3,warm4}.}. To complete our discussion of inflationary scenarios, we would like to adopt a different (but complementary) perspective of showing that no modifications of general relativity are really required to explain the Planck results and that, in fact, there is a degeneracy between the predictions of the Starobinsky model and that of generic plateau-like potentials within general relativity with a scalar inflaton.

\begin{figure}[htbp]  
 \centering\includegraphics[scale=1]{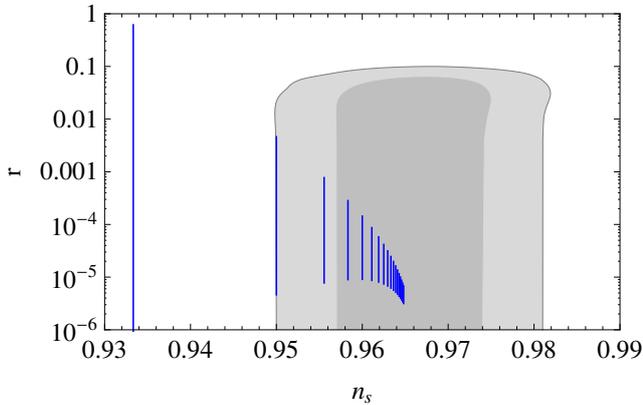}
 \caption{Observational predictions in the ($n_s$,$r$) plane for plateaux of order $p=3, \ldots, 20$ (increasing in unit steps from left to right), for 60 e-folds of inflation and $\mu=10^{-3}-1$. Planck data is shown in grey.}
 \label{fig:plateau}
 \end{figure}

For this, we define a $p$-plateau, $p>2$, as a potential function for which $p-1$ derivatives vanish at some field value $\phi_0$, with $V^{(p)}(\phi_0)<0$. Close to the central plateau value $\phi_0$, the potential may thus be written in the form:
\begin{equation}
V(\phi)= V_0\left[1-{\mu\over p}\left({\phi-\phi_0\over M_P}\right)^p+\ldots\right]~,
\end{equation}
where $\mu>0$. Defining $\Delta\tilde\phi = (\phi-\phi_0)/M_P\ll 1$, one has for the slow-roll parameters:
\begin{eqnarray}
\epsilon_\phi&\simeq& {\mu^2\over2}\Delta\tilde\phi^{2p-2}~, \nonumber\\
\eta_\phi &\simeq &-\mu (p-1) \Delta\tilde\phi^{p-2}~,
\end{eqnarray}
such that $\epsilon_\phi\ll |\eta_\phi|$ for $\mu\lesssim 1$. The number of e-folds of inflation after horizon-crossing of the relevant CMB scales is thus:
\begin{eqnarray}
N_e= -{1\over M_P^2}\int_{\phi_*}^{\phi_e} {V(\phi)\over V'(\phi)}d\phi \simeq  {\Delta\tilde\phi_*^{2-p}\over (p-2)\mu}~,
\end{eqnarray}
where we have taken into account that $\Delta\tilde\phi_e\gg \Delta\tilde\phi_*$ and $p>2$, while the spectral index and tensor-to-scalar ratio are given by:
\begin{eqnarray}
n_s-1 &=& -6\epsilon_{\phi_*}+2\eta_{\phi_*} \simeq -2\mu (p-1) \Delta\tilde\phi_*^{p-2} \nonumber\\
&\simeq& -{p-1\over p-2}{2\over N_e}~,\nonumber\\
r &=& 16\epsilon_{\phi_*}\simeq 8\mu^2\Delta\tilde\phi_*^{2p-2} \simeq 8\mu^{s}(p-2)^{s-2} N_e^{s-2}~,\nonumber\\
\end{eqnarray}
where $s=2/(2-p)$. For $p\gg 1$, i.e.~for a very flat plateau, one then has:
\begin{eqnarray}
n_s&\simeq &1-{2\over N_e}~,\nonumber\\
r&\simeq & {8\over p^2 N_e^2}~,
\end{eqnarray}
 where the scalar spectral index prediction coincides with that of the Starobinsky potential and Higgs-inflation with a non-minimal gravitational coupling, yielding $n_s\simeq 0.96-0.967$ for 50-60 e-folds of inflation. We show in Fig.~\ref{fig:plateau} the predictions for different plateaux with integer $p\geq 3$ and $\mu=10^{-3}-1$ for 60 e-folds of inflation after horizon-crossing. It is clear from this figure that observational predictions become insensitive to the curvature of the potential, $\mu$, for large $p$, and that to agree with Planck data at 95\% C.L.~requires a flat plateau of integer degree $p\geq6$. 
  
 Note that the tensor-to-scalar ratio can be arbitrarily large for any finite value of $p$ by taking $\mu\to0$, and in this sense such plateaux potentials are degenerate with the Starobinksy model, for which $r\simeq 12/N_e^2$. The effective scalar potential in this model, $V(\phi)= V_0\left(1-e^{-\sqrt{2/3}\phi/M_P}\right)^2$, may in fact be seen as an infinite degree plateau at $\phi_0\rightarrow +\infty$ (for which the above analysis does not strictly apply). This nevertheless shows that Planck is not really pointing definitely towards a modification of gravity, but simply to any model that effectively leads to a sufficiently flat plateau in the scalar potential, within a general relativistic description.

\section{Conclusions}\label{sec:conclusions}
Inflation is an extremely successful paradigm that solves the main problems of the Hot Big Bang model, also providing an explanation for the origin of the cosmic structures that we observe today. There are several models that are compatible with such a mechanism, and these are very constrained or even ruled out by recent precise data obtained by the Planck mission \cite{planck}.

In this work we have analysed some examples of inflationary models that are based on different modifications of general relativity and the gravitational interactions of the scalar inflaton field. 

For the inflationary model inspired by the generalised Chaplygin gas, which unifies dark matter and dark energy at late times, we have obtained agreement with the Planck data for a linear potential and for quadratic and quartic hilltop potentials, with the quadratic monomial potential being ruled out. We have also found that the departure from general relativity leads to an attractor at $(n_s,r)=(1, 0)$ as the parameter $x = A/V _0^{1+\alpha}$ grows.

We have also obtained similar agreement with data for inflationary scenarios on a 3-brane for a linear potential and both the quadratic and quartic hilltop potentials, with larger deviations from general relativity leading to smaller values of the tensor-to-scalar ratio. Again the quadratic monomial model is ruled out in the brane scenario.

We have also considered an inflationary model within $N=1$ supergravity with an $R$-symmetry, yielding a hilltop-like potential with vanishing first and second derivatives at the origin, which is observationally disfavoured at more than $2\sigma$. A similar model can, however, be made compatible with the Planck data by considering a brane-localised scenario.

Finally, we have pointed out that, although we have found agreement with data for several inflationary models within modified gravity, observations do not really require going beyond general relativity, being sufficient to consider, within the context of the latter, a scalar potential with a very flat plateau section. Predictions for such models can, in fact, be degenerate with those of the popular Starobinsky model, which is essentially likewise a special plateau-potential in the Einstein frame.

Breaking this degeneracy will require not only better precision measurements of the scalar spectral index, but also of other quantities, namely those related to the primordial tensor spectrum and possibly higher-order corrections to power-law spectra such as running indices. For example, similarly to what we have obtained in Ref. \cite{nmcinf}, inflationary models with a modified Friedmann equation such as the Chaplygin-inspired model or brane inflation analysed in this work will lead to a modified consistency relation between the tensor spectral index and the tensor-to-scalar ratio, which can be used to assess whether modifications of the gravitational sector play an important role in the inflationary dynamics. We hope that this work motivates further exploration of such observables and consistency relations within different models in order to better understand the type of novel physics required to successfully realise the inflationary paradigm.


\section*{Acknowledgments}

The work of C.G. is supported by Fundação para a Ciência e a Tecnologia (FCT)
under the grant SFRH/BD/102820/2014. J.G.R. is supported by the FCT Investigator Grant No.~IF/01597/2015 and partially by the H2020-MSCA-RISE-2015 Grant No. StronGrHEP-690904 and by the CIDMA Project No.~UID/MAT/04106/2013.



\end{document}